\documentstyle[12pt]{article}
\newcommand{\be}{\begin{equation}}
\newcommand{\ee}{\end{equation}}
\title{\bf
Generalized coordinate gauge and nonabelian
 Stokes theorem}
\author{ V.I.Shevchenko\thanks{shevchen@heron.itep.ru} and 
Yu.A.Simonov\thanks{simonov@vxitep.itep.ru}.
\\
{\it Institute for Theoretical and Experimental Physics}\\
{\it 117218, B.Cheremushkinskaya, 25, Moscow, Russia}}
\date{}
\begin{document}
\maketitle
\vspace{1cm}
\centerline{\bf {Abstract}}
\vspace{3mm}
A contour gauge of general type is analysed 
where 1-form (vector potential) is expressed as a contour integral of the 
2-form (field strength) along an arbitrary contour $C$. 
For a special class of contours the gauge condition reduces to
$k_{\mu}(x) A_{\mu}(x) = 0 $ where $k_{\mu}(x)$ is a tangent vector to
the contour $C$. It is shown, that this gauge
is advantageous to give a simple proof of the nonabelian Stokes theorem.

\newpage

 The procedure of the gauge fixing is 
an essential part of QCD \cite{ynd} and however final results do not
depend on the gauge, different forms of gauge conditions 
are useful in different settings of physical problems. 
For example, in high energy scattering in QCD the axial 
gauge has proved to be useful \cite{lip}, while in the OPE analysis 
\cite{sh} the Fock-Schwinger \cite{fss} (sometimes called the 
coordinate or radial) gauge was applied (for
discussions and derivation see  \cite{sh1} and also \cite{dub}).

In another physical situation, where the time axis is 
singled out, as e.g. in the heavy quarkonium theory,
the modified coordinate gauge \cite{bal} can be convenient. 
This  gauge was used recently in the context of equations 
for the quark \cite{sim1} and gluon \cite{sim2} Green's functions,
displaying the property of chiral symmetry breaking and confinement.

There is another set of studies where an emphasis is made on 
formulation of gauge theory without gauge-dependent degrees
of freedom from the very beginning, and the role of dynamical
variables is played by 2-forms \cite{mand} or loop variables \cite{mm}.
These completely gauge-invariant approaches encountered their own
difficulties and as a matter of fact many gauge-invariant observables
are easier to calculate using gauge-dependent diagrammatic
rules.

Both in the coordinate gauge \cite{fss} and in its modified form
\cite{bal} the shape of the contour $C(x)$, in the integral, 
connecting vector potential and the field strength,
\be
A_{\mu}(x) = \int\limits_{C(x)} dz_{\nu} {\alpha}_{\rho\mu}(z) 
F_{\nu\rho}(z)
\label{stoks1}
\ee
is fixed and consists of straight lines. Inessential for
physical results, it may be inconvenient in the course
of computations. In particular, in the confining phase of QCD,
when the QCD string is formed between two colour charges 
it would be advantageous to choose the contours $C$ lying on the 
world sheet of the string; in this case one could do simplifying
approximations as in \cite{sim1,sim2}, namely to keep only
Gaussian field correlator. The decoupling of ghosts,
known to occur for the gauges (\ref{stoks1})
(see \cite{leo}, \cite{kor} and references therein) is also an attractive 
feature, which suggests to look for generalizations of 
(\ref{stoks1}) with arbitrary contours $C$.

The gauge condition of the type we are
interested in was introduced for the first time in \cite{kor}.
In the present paper we give a refined treatment of this gauge,
paying special attention to some important
details, missing in the original paper. 
Let us briefly mention them.
To define this gauge condition correctly, 
the set of contours $C$, determining the gauge must satisfy some
additional requirement (eq.(\ref{auto}) of the present paper).
This condition is essential for the representation (\ref{stoks1})
to hold true.
With this requirement we are also able to 
formulate
the gauge condition in the local form (eq.(\ref{gauge}) of
the present paper).  
An immediate use of the generalized contour gauge 
which also has not yet been discussed in the literature 
is the ability
to give a short and direct proof of the nonabelian Stokes 
theorem \cite{halpern,aref} as we do below in this paper.

Let us proceed with the definition of the generalized contour gauge.
Let $M$ be a $d$-dimensional connected Euclidean manifold.
We choose some subspace $M_0$, $M_0 \subset M$
which
in general may be disconnected and of lower dimension than $d$.
For our purposes it is sufficient 
to take $M_0$ consisting of the only one point $x_0$.
If $M_0$ is of more complicated structure  
some specific features appear, which we
plan to discuss 
elsewhere (see e.g. the remark before the eq.(\ref{qq})).

For each point $x \in M\setminus  M_0$ we define the unique smooth contour
$C_{x_0}^{x}$, $x_0 \in M_{0}$ connecting points $x$ and $x_0$.
The contours are parametrized as follows:
\be
C_{x_0}^{x} :\;  z_{\mu} = z_{\mu}(s,x);
\> s\in[0,1];\> z_{\mu}(0,x) = {x_0}_{\mu};\>
z_{\mu}(1,x) = x_{\mu}
\ee
The map $M\setminus  M_0 \to M_0$ defined above is
naturally extended to $M \to M_0$ by setting 
$C_{x_0}^{x_0}$ to be the unit contour: $z_{\mu}(s,x_0)\equiv {x_0}_{\mu}$.
The resulting map $M \to M_0$ is assumed to be smooth.
In the particular case when $M_0$ consists of the only point $x_0$
it means, that the manifold $M$ should be contractible.

Let us choose two arbitrary points
$z_{\mu}(s,x)$ and $z_{\mu}(s',x)$
on the given contour $C$
in such a way that the point $z_{\mu}(s',x)$
lies between points $z_{\mu}(s,x)$ and $z_{\mu}(1,x)=x_{\mu}$ 
(if $s$ is natural parameter, it simply 
means that $s<s'$). We 
assume the following condition - for any $s,s'$ there exists
$s''$ such that 
\be
z_{\mu}(s,x) = z_{\mu}(s'',z(s',x))
\label{auto}
\ee
The geometrical meaning of (\ref{auto}) is simple:
for any point $z$ lying on some contour $C_{x_0}^{x}$
its own contour $C_{x_0}^{z}$ coincides with the corresponding part 
of the contour $C_{x_0}^{x}$. 
The eq.(\ref{auto}) does not mean, generally speaking, that 
contours 
$C_{x_0}^{x_1}$ and $C_{x_0}^{x_2}$
from different points $x_1 \neq x_2$ 
have no common points except $x_0$.
The condition that contours $C_{x_0}^{x}$ should not selfintersect
(the only condition discussed in \cite{kor}) is 
necessary but not sufficient 
to guarantee (\ref{auto}) and therefore to derive (\ref{rota})
and (\ref{gauge}) below.
The defined set of contours forms an oriented tree graph without
closed cycles according to (\ref{auto}).

Let us now start with the gauge potential $A_{\mu}(x)$ taken in some
arbitrary gauge and perform the gauge rotation
\be
A_{\mu}'(x) = {\Omega}^{+}(x) A_{\mu}(x) {\Omega}(x) + \frac{i}{g}
\> {\Omega}^{+}(x) {\partial}_{\mu} {\Omega}(x)
\label{ga}
\ee
with 
$$
{\Omega}(x) = U(x,x_0) = Pexp{(ig\int\limits_{x_0}^{x} A_{\mu}(z) dz_{\mu})}
$$
and integration goes along the contour $C_{x_0}^x$.  
The important point is the differentiation of the phase factors
\cite{mand,mm} which is a well defined procedure for our choice 
of contours since the function $z_{\mu}(s,x)$ is given:
$$
{\partial}_{\mu} {\Omega}(x) = ig A_{\mu}(x) {\Omega}(x) +
$$
\be
+ ig \int\limits_0^1 ds \frac{\partial z_{\nu}(s,x)}{\partial s} 
{\alpha}_{\rho\mu}(z) U(x,z(s)) F_{\rho\nu}(z(s))
U(z(s),x_0)
\label{dif}
\ee
where 
\be
{\alpha}_{\rho\mu}(z) = \frac{\partial z_{\rho}(s,x)}{\partial x_{\mu}}
\ee

Substituting (\ref{dif}) into the (\ref{ga}) one gets:
\be
A_{\mu}'(x) = - U(x_0,x)
\int\limits_0^1 ds \frac{\partial z_{\nu}(s,x)}{\partial s} 
{\alpha}_{\rho\mu}(z) U(x,z(s)) F_{\rho\nu}(z(s))
U(z(s),x_0)
\label{kro}
\ee
Taking into account the condition (\ref{auto})
and  the gauge transformation property
$$
U(x_0,x)U(x,z(s)) F_{\rho\nu}(z(s))
U(z(s),x_0) \to {F}_{\rho\nu}'(z(s))
$$
we arrive to the final result (omitting primes in what
follows)
\be
A_{\mu}(x) =  
\int\limits_0^1 ds \frac{\partial z_{\nu}(s,x)}{\partial s} 
\> \frac{\partial z_{\rho}(s,x)}{\partial x_{\mu}}
\> F_{\nu\rho}(z(s))
\label{rota}
\ee
This formula was proposed in \cite{kor} and used without 
derivation in \cite{sim2}.
Note at the same time, that the condition (\ref{auto}), is crucial 
for proceeding from (\ref{kro}) to (\ref{rota}).

The eq.(\ref{rota}) leads to important local
condition for vector-potential.
To this end, note, that 
solving (\ref{auto})
with respect to $s'$ we find:
$$
s' = f(s,s'',x);\;\; f(s,s,x) = 1
$$
Substituting $s'=f(s,s'',x)$ into (\ref{auto}) and differentiating 
with respect to $s$ one gets:
\be
\frac{\partial z_{\mu}(s,x)}{\partial s} = 
\frac{\partial z_{\mu}(s'',z(s',x))}{\partial 
z_{\rho}(s',x)}\frac{\partial z_{\rho}(s',x)}{\partial s'}
\frac{\partial f(s,s'',x)}{\partial s}
\label{didi}
\ee
By putting $s'$ to be equal to unity (\ref{didi}) reads:
\be
\frac{\partial z_{\mu}(s,x)}{\partial s} = 
\frac{\partial z_{\mu}(s,x))}{\partial 
z_{\rho}(1,x)}\left.\left(\frac{\partial z_{\rho}(s',x)}{\partial s'}
\right)\right|_{s'=1}\cdot g(s,x)
\label{ff}
\ee
with $g(s,x) = (\partial f(s,s'',x) / \partial s )_{s''= s}$.
We can now multiply both sides of (\ref{rota}) by 
$k_{\mu}(x) = (\partial z_{\mu}(s,x)/\partial s)_{s=1}$ and get 
$$
A_{\mu}(x)\cdot k_{\mu}(x) = 
$$
$$
=  
\int\limits_0^1 ds \frac{\partial z_{\nu}(s,x)}{\partial s} 
\> \frac{\partial z_{\rho}(s,x)}{\partial x_{\mu}}
\left.\left(\frac{\partial z_{\mu}(s,x)}{\partial s}\right)\right|_{s=1} 
\> F_{\nu\rho}(z(s)) =
$$
\be 
= \int\limits_0^1 ds \frac{\partial z_{\nu}(s,x)}{\partial s} 
\> \frac{\partial z_{\rho}(s,x)}{\partial s}\> {(g(s,x))}^{-1}
\> F_{\nu\rho}(z(s)) = 0
\label{ort}
\ee
where the second equality holds by virtue of (\ref{ff})
and the third due to antisymmetry of $F_{\rho\nu}$.
The condition (\ref{ort}) can be easily understood 
taking into account, that phase factors along the contours $C_{x_0}^{x}$ specifying
the gauge are equal to unity:
$$
U(x,x_0) = Pexp{(ig\int\limits_{x_0}^{x} A_{\mu}(z) dz_{\mu})} =1
$$
Since (\ref{ort}) holds for any $x$, one gets:
\be
A_{\mu}(x) \> k_{\mu}(x) = 0
\label{gauge}
\ee 

Specific examples of the gauge condition discussed in the present
article are known in the literature. 
The best studied is 
the radial or Fock-Schwinger gauge
\cite{fss}. 
In this gauge the set of contours is defined by 
\be 
z_{\mu}(s,x) = s\cdot x_{\mu}
\label{gnio}
\ee 
and (\ref{gauge}) reads:
\be
x_{\mu}A_{\mu}(x) =0
\ee
The relation (\ref{rota}) becomes 
\be
A_{\mu}(x) =  
\int\limits_0^1 ds  s x_{\nu}
\> F_{\nu\mu}(z(s))
\ee
Note that due to the topological restrictions 
stated above this gauge condition 
is already defined in some neighbourhood of the point $x_0=0$
but might not be well 
defined globally. This was noticed in different respect 
also in {\cite{leo}}. 

Another example is the gauge condition introduced in \cite{bal}.
It singles out not a point as the Fock-Schwinger gauge but a line.
One can take this line to be defined by $z_1=z_2=z_3=0$. 
Then the contours $C(x)$ are made of two straight paths: 
$$ 
z_i(s,x) = q(s) \cdot x_i, \> i=1,2,3;\; z_4(s,x) = x_4 + p(s,x)\cdot n_4
$$
where $n_4$ is the unit vector in the forth direction,
$q(s)$ is the linear function, satisfying $q(1)=1,\> q(s_0) = 0$ 
and the function $p(s,x)$ is such that  $p(s,x)\equiv 0$ for 
$1\ge s\ge s_0$ and $p(s,x)\to \infty$ if $s\to 0$.
The choice of $s_0$ is arbitrary.

Piecewise nature of $C$ leads to two different conditions 
depending on whether
the point $x$ lies on the singled out line or not.
By imposing an additional requirement 
$$
\left.(\partial p(s,x) / \partial s)\right|_{s=s_0} 
= 1
$$
we obtain from (\ref{gauge}):
$$
A_4(x_i=0) = 0;\; A_i(x)\cdot x_i = 0
$$
This gauge was proved to be useful in the 
studies of heavy quarkonium dynamics \cite{bal,vairo}.

A natural generalization of the above conditions is a gauge,
which could be called planar and it was actually used without
derivation in \cite{sim2}.
It is constructed by choosing a plane $z_1=z_2=0$ and the contours
$C$ to be orthogonal to this plane.
Then (\ref{gauge}) reads:
$$
A_1(x) x_1 + A_2(x) x_2 = 0
$$
and additional gauge freedom for the potentials
on the plane itself still remains to be fixed in an appropriate way. 
  
It should be obvious from the derivation and considered examples that
the gauge conditions of the type (\ref{gauge}) are generally not
enough to fix the gauge modulo global transformations
(see discussion of this point for the radial gauge in \cite{leo}). 
But if the base manifold $M$
satisfies some requirements (contractible if $M_0$ consists of the only point,
as it is, for example, in the case of radial gauge) 
the additional gauge freedom is absent. We plan to discuss these 
questions in detail in subsequent publication.

As an illustrative example let us consider the use of the
generalized gauge
condition for
the nonabelian Stokes theorem. 
There are different proofs of this
theorem in the literature 
\cite{halpern,aref}, but what
we are going to present is perhaps the simplest one.
It is close in spirit to the 
paper \cite{halpern}. Namely, we define the gauge condition 
in such a way that potential $A_{\mu}(x)$ on the contour is 
expressed as a
function of field strength $F_{\mu\nu}(u)$ defined on the (arbitrary) 
surface, bound by the contour. Then rewriting gauge-invariant Wilson
loop in this gauge we obtain a relation, valid in the chosen
gauge and as the last step put it into gauge-covariant form. 
It was done in \cite{halpern} for the completely
fixed axial gauge condition, which is a convenient choice  
in two dimensions (or for planar surfaces in higher 
dimensional case). Our procedure allows one to 
choose an arbitrary surface $S$ bound by the 
simple contour $C = \partial S$ and therefore the gauge
condition we use entirely depends on the shape of $S$.

We parametrize the surface
 by $w_{\mu}(s,t);\> s,t \in [0,1] $.
We choose an arbitrary point ${x_0}_{\mu}$ on the surface
 in such a way, that
$w_{\mu}(0,t) \equiv {x_0}_{\mu}$.
If $s=1$ then  $w_{\mu}(1,t)$ goes along the contour $C$
and $w_{\mu}(1,0) = w_{\mu}(1,1)$ according to $\partial C = 0$.

The following important remark is in order.
It is usually assumed that $S$ has the disk topology, in this
simplest case we are free in our choice of $M_0$, which may
consist of only one point, what we actually have used. 
For this topology 
it is always possible to define a set of contours obeying 
(\ref{auto}) by continuous
deformation of the planar disk with the radial contours (\ref{gnio}).
Indeed, the continuous, without cuts and gluings, 
deformation of the surface 
leaves (\ref{auto})
intact for the contours, defined on this surface.
But if $S$ is one-hole surface of nontrivial genus,
the proof should be modified, because 
in this case $S$ cannot be retracted to a point.
In other words, it is impossible to define the smooth set
of contours, obeying (\ref{auto}) on a higher genus surface   
if $M_0$ consists of only one point.
Instead $M_0$ must be taken as 1-cycle. 
So the validity of the nonabelian 
Stokes theorem in this case depends on the
possibility to retract arbitrary surface with a hole to $S^1$. 
We plan to discuss the subtleties of 
nonabelian Stokes theorem for the surfaces
of higher genus in subsequent publication.

According to (\ref{rota}) the gauge potential $A_{\mu}(z)$ is related to
$F_{\mu\nu}(z)$ in the following way:
\be
A_{\mu}(z(s,t))\>=\>
 \int\limits_0^1 ds' \frac{\partial z_{\nu}(s',x(t))}{\partial s'} 
\> \frac{\partial z_{\rho}(s',x(t))}{\partial x_{\mu}(t)}
\> F_{\nu\rho}(z(s',t))
\label{qq}
\ee
Equation (\ref{qq}) is actually nothing else than the Stokes theorem in its 
infinitesimal form. It is well known that the generalization to  
finite contours is nontrivial in the nonabelian case,
in particular the integral $\int_S F_{\mu\nu} d\sigma_{\mu\nu} $ depends on the
surface even if the contour $C=\partial S$ is closed.
But this integral does not enter by itself in the 
nonabelian Stokes theorem. Instead
the quantity which should be considered here is a $P$-ordered
exponent, by definition it reads:
$$
Pe^{ig\int\limits_{C} A_{\mu}(x) dx^{\mu} } = \hat 1 +
$$
\be
+ \sum\limits_{n=1}^{\infty} {(ig)^n}
\int ..\int 
d{x^{(1)}_{\mu 1}}.. d{x^{(n)}_{\mu n}}\> 
{A_{\mu}}_n(x^{(n)})..{A_{\mu}}_1(x^{(1)})
\>\theta (x^{(1)} > x^{(2)} >..> x^{(n)})
\label{st}
\ee
The $\theta$-function in (\ref{st}) orders the points $x^{(i)}$ along 
the contour $C$. 

Substitution of (\ref{qq}) into (\ref{st}) leads to the expression:
$$
Pe^{ig\int\limits_{C} A_{\mu}(x) dx_{\mu} } =
$$
$$
=  1 +  \sum\limits_{n=1}^{\infty} {(ig)^n}
\int ..\int d\sigma_{\mu\nu}(w^{(1)}(s_1,t_1)).. d\sigma_{\rho\phi}
(w^{(n)}(s_n,t_n))\> 
$$
\be
 F_{\rho\phi}(w^{(n)}(s_n,t_n)).. F_{\mu\nu}(w^{(1)}(s_1,t_1))
\>\theta \left(t_1 > t_2 >..> t_n \right)
\label{sst}
\ee
Note, that ordering procedure in (\ref{sst}) is the same as for
the original Wilson loop -- only the points along the contour $C$ are
ordered with respect one to another, i.e. 
ordered in parameters $t_i$, while the 
integrals over $s_i$ are taken independently for each $t_i$.

To bring (\ref{sst}) to the gauge covariant form
 we introduce phase factors along the $s$-direction on the surface,
 which are equal to unity due to (\ref{gauge}), 
 i.e. we replace
$$
F_{\mu\nu}(w(s,t)) \to G_{\mu\nu}(w(s,t)) = 
U(x_0; w(s,t))F_{\mu\nu} (w(s,t)) U(w(s,t); x_0)
$$
Under arbitrary gauge rotations the l.h.s. of (\ref{st})
transforms as:
\be
Pe^{ig\int\limits_{C_{x^{*}x^{*}}} A_{\mu}(x) dx_{\mu} } \to
\Omega^{+}(x^{*})
Pe^{ig\int\limits_{C_{x^{*}x^{*}}} A_{\mu}(x) dx_{\mu} } 
\Omega(x^{*})
\ee
where $x^{*}$ is an arbitrary fixed point on the contour $C$
(lower limit in all integrals in (\ref{st})).
If the point $x_0$ does not lie on the contour $C$, then
the final gauge-covariant answer reads:
\be
 Pe^{ig\int\limits_{{C}_{x^{*}x^{*}}} A_{\mu} dz_{\mu} } =
U(x^{*}, x_0) 
{\cal P} e^{ig \int\limits_{S} d\sigma_{\mu\nu}(z) G_{\mu\nu}(z)}
U(x_0,x^{*})
\label{bre}
\ee
where the meaning of the ordering simbol $\cal P$ is explained
in (\ref{sst}).
Under the gauge rotations both sides of (\ref{bre}) are transformed 
in the same way which finishes the proof.
The more often used gauge-invariant form of (\ref{bre}) is
\be
Tr\> Pe^{ig\int\limits_{C} A_{\mu} dz_{\mu} } =
Tr\> {\cal P} e^{ig \int\limits_{S} 
d\sigma_{\mu\nu}(z) G_{\mu\nu}(z)}
\label{qsi}
\ee
We stress again, that the exact meaning of the symbol ${\cal P}$ 
is completely determined by the choice of the set of contours, 
defining the gauge which may be done as the most convenient one 
for a given application of the nonabelian Stokes theorem.

The advantages of the discussed gauge condition 
are not 
exhausted by the simple and transparent proof of Stokes theorem
given above, in particular see \cite{kor} for some applications. 
In our opinion, the most interesting development 
has not yet been investigated.
Namely, in the partition function one could introduce the 
integration over the set of contours
defining the gauge (\ref{rota}) in addition to the 
integration over field strengths. 
This contour integration is natural to associate with the 
sum over the surfaces,
bounded by the Wilson contour $C$. This 
yields a choice of integration variables 
alternative to what is 
usually discussed
in the literature \cite{polik}.
We plan to develop this issue in future publications.
\bigskip

{\large\bf Acknowledgments}

The work was supported by the RFFI grants 96-02-19184a, 96-15-96740,
and partially by RFFI-DFG grant 96-02-00088G.
The authors are grateful to A.Levin, M.Olshanetsky, V.Novikov 
for valuable discussions and to G.Korchemsky for pointing authors
attention to the Ref.\cite{kor}.

\vspace{3mm}


\vspace{6mm}


\begin{thebibliography}{99}
\bibitem{ynd}
L.D.~Faddeev, "Gauge Fields: An Intro to Quantum Theory",
Addison-Wesley, 1991.\\
F.~Yndurain, "The Theory of Quark and Gluon Interactions", 2-nd 
ed., Springer-Verlag, New-York, 1993.
\bibitem{lip}
L.V.~Gribov, E.V.~Levin, M.G.~Ryskin, Phys.Rep., {\bf 100} (1983) 1. 
\bibitem{sh}
V.~Novikov, M.~Shifman, A.~Vainshtein, V.~Zakharov,
Nucl.Phys.{\bf B249} 445 (1985).
\bibitem{fss}
V.A.~Fock, Sov.Phys.,{\bf 12} (1937) 404.\\
J.~Schwinger, Phys.Rev.,{\bf 82} (1952) 684, see
also  "Particles, Sources and Fields", 
vols.1 and 2, Addison-Wesley, Reading, 1970 and 1973.
\bibitem{sh1}
M.~Shifman, Nucl.Phys.{\bf B173} 13 (1980).
\bibitem{dub}
M.S.~Dubovikov, A.V.~Smilga, Nucl.Phys., 
{\bf B 185} (1981) 109.
\bibitem{bal}
I.Balitskii, Nucl.Phys.{\bf B 254} (1985) 166.
\bibitem{sim1}
Yu.A.~Simonov, hep-ph/9704301 \\
Yu.A.~Simonov, hep-th/9712250
\bibitem{sim2}
Yu.A.~Simonov, hep-ph/9712248
\bibitem{mand}
S.Mandelstam, Phys.Rev. {bf 175} (1968) 1580;\\
Ann.Phys. {\bf 19}, 1 (1962) 25. 
\bibitem{mm}
Yu.M.~Makeenko, A.A.~Migdal, Nucl.Phys.{\bf B188} (1981) 269,\\
for a review see A.A.~Migdal, Phys.Rep.{\bf 102} (1983) 199.
\bibitem{leo}
S.~Leupold, H.~Weigert, hep-th/9604015.
\bibitem{kor}
S.V.~Ivanov, G.P.~Korchemsky, Phys.Lett. {\bf B154} (1985) 197.\\
S.V.~Ivanov, G.P.~Korchemsky, A.V.~Radyushkin,
Sov.J.Nucl.Phys. {\bf 44} (1986) 145.
\bibitem{halpern}
M.~Halpern, Phys.Rev.{\bf D19} (1979) 517.
\bibitem{aref}
Y.~Aref'eva, Theor.Math.Phys. {\bf 43} (1980) 353;\\
N.Bralic, Phys.Rev. {\bf D22} (1980) 3090;\\
Yu.A.~Simonov, Sov.J.Nucl.Phys. {\bf 50} (1989) 134;\\
M.Hirayama, S.Matsubara, hep-th/9712120. 
\bibitem{vairo}
N.~Brambilla, A.~Vairo, Phys.Lett. {\bf B 407} (1997) 167.
\bibitem{polik}
E.~Akhmedov, M.~Chernodub, M.~Polykarpov, M.~Zubkov,\\
Phys.Rev. {\bf D 53} (1996) 20.\\
A.M.~Polyakov, Nucl.Phys., {\bf B486} (1997) 23.

\end{thebibliography}
\end{document}